# Measurement of Global Value Chain (GVC) Participation in World Development Report 2020

Sourish Dutta

PhD Student

Centre for Development Studies

Trivandrum, Kerala

**Abstract**

As we can understand with the spread of GVCs, a lot of new questions emerge regarding the measurement of participation and positioning in the globalised production process. The World Development Report (WDR) 2020 explains the GVC phenomenon and then focus on participation and the prospects especially in a world of change in technology. From the overview section, we can figure out that nowadays, goods and services flow across borders as intermediate inputs rather than final goods. In traditional trade, we need two countries with the notions of export and import. However, in GVC trade, the goods and services cross borders multiple times requiring more than two countries. Remarkable improvements in information, communication, and transport technologies have made it possible to fragment production across national boundaries. So the question is: how to conceptualise this type of new trade to justify the measurement of participation.

**Conceptualisation of GVC**

Strengthening of trade liberalisation with continuing reduction of transportation cost, revolution in the information and communications technology (ICT), and some recent policy reforms are expanding the scope of globalisation (Antras, 2015). Indeed it is forming a complex network called 'Global Value Chains' (GVCs) through the distribution of designing, procuring parts & components, and assembly across several countries' firms in different sectors. In a proper sense, GVC is a series of production stages of goods & services adding value from initial conception to final consumption with at least two stages in different countries. It can take spider-like (convergence of parts) and snake-like (sequence of parts) compositions (Baldwin and Venables, 2013) with the task-level division of labour having great productive use of resources across countries, sectors, and firms. For example, bicycles go through designing, prototyping, and conception work in Italy, producing parts and components in Italy, China, Japan, Malaysia, and many countries, and then assembling of all inputs in Taiwan and China. This efficient organisation of bicycle value chain using different parts and components around the world results in a low-cost & high-quality bicycle to the consumer (Kalm et al., 2013; OECD, 2013). This type of globalisation can stimulate industrialisation in developing countries by giving access to a broad market, cheaper & better inputs, technologies, and management practices. However, GVC is not wholly a new thing because a significant proportion of world trade in raw materials and intermediate inputs has been hinting at this phenomenon. Although the growth of GVCs with augmented global trade & investment outpacing economic output increased from 1990 to 2008, the 2008 financial crisis and its consequent recessionary conditions disturbed its organisation, causing cyclical and structural economic uncertainties.

**Measuring GVC Participation**

We can understand that with the spread of GVCs, a lot of new questions arise about the measurement of participation and positioning in the globalised production process. From the earlier section, we can figure out that nowadays goods and services flow across borders as intermediate inputs rather than the final goods. In traditional trade, we need two countries with the notions of export and import. However, in GVC trade, the goods and services cross borders multiple times requiring more than two countries. So the question is: how to conceptualise this type of new trade to justify the measurement of participation & position. There is a concept called 'import to export' (Baldwin & Lopez-Gonzalez, 2015), i.e. use of imported content (in terms of value-added embodied in materials, intermediate inputs or tasks) for exports at the country-industry-firm level. There is one difficulty in using this notion to measure GVC participation. Because the well-known customs data to measure standard trade is not enough to answer the questions of the GVC trade phenomenon regarding how the exported goods and services are produced and how it will be used in importing country. Hence the issue of measurement is that the traditional trade statistics do not represent an adequate picture of supply and demand linkages among the economies.

We need new data, new methods, and new tools for a full description of a country's exposure to world demand. Here the new data are global input-output tables (involving customs data and national input-output tables). The national input-output (IO) tables are the production and consumption structures within an economy. If we generalise these national IO tables to describe the sale and purchase relationships between producers and consumers within and between economies, then we get the global or inter-country IO (ICIO) tables. In a national IO table, exports are sales to the 'foreign sector'; in an ICIO table, exports are to country 1, county 2, and like this. The most widely used ICIO tables are OECD-Trade in Value Added (TiVA), World Input-Output Database (WIOD), EORA-MRIO. World Development Report 2020 has used the EORA data because of its most substantial country coverage for the most prolonged period. However, its sectoral coverage is more aggregate and so less accurate than the WIOD and TiVA databases (Lenzen et al., 2012; Borin and Mancini, 2019; Johnson, 2018). The new methods are related to a better understanding of value-added and GVC trade at aggregate, bilateral and bilateral-sectoral level trade. These methods answer how to measure VA content at the aggregate, bilateral and bilateral-sectoral level, depending on the specific empirical issue developing a measure of GVC-related trade reconciling large part of the existing literature under one framework. The new tools are the R and Stata programs for economic analysis with ICIO tables.

In this context of measuring GVC trade (Borin and Mancini, 2019) in total world trade (as a share of trade that flows through at least two borders), usually, two measures are used: backward and forward GVC participation. Backward GVC participation shows a country's exports embody value added that it has previously imported from abroad. Whereas, forward GVC participation shows a country's exports are not fully absorbed in the importing country, and instead are embodied in the importing country's exports to third countries. In other words, GVC participation is termed as 'backward' if the intermediate inputs are from the preceding stage of production and termed 'forward' if the exporter is at the early stage of production. For example, India exports aluminium tubing to Taiwan and China, where it is further used in the production of the bicycle later exported. Here India's participation is considered as forward, and China & Taiwan participate in a backward way. These computations are based on the global input-output tables with broadly defined sectors (scaled-up versions of product-level studies like the bicycle) and strong assumptions regarding the flow of intermediate inputs (De Gortari, 2019). In practice, the GVC phenomenon is all about firm-level international trade. This firm-level phenomenon raises some new issues concerning the heterogeneity in terms of productivity across firms and intrafirm trading hints about different types of buyer-seller relations (Johnson, 2018).

**Evolution of GVC Participation**

The overall percentage of GVC trade in the whole international trade (comprising both forward & backward participation) increased quite significantly from 1990 to 2008. Although it seems to have stagnated or even dwindled in the last decade, yet about half of international trade looks to be linked to global value chains. In the case of India, the share of GVC trade improved from 25 per cent to around 35 per cent during the same period. The forward participation mainly dominates it. Nevertheless, the growth of backward participation rose more from around 8 per cent to 16 per cent.

The WDR 2020 has sequentially classified four types of GVC participation (considering country sizes) on the basis of (1) sectoral specialisation of exports (based on domestic value added in gross exports of primary goods, manufacturing, and business services); (2) extent of GVC participation (measured as backward integration of the manufacturing sector as share of the country's total exports, i.e. backward manufacturing), where higher backward integration in manufacturing is an essential characteristic of countries entering or specialised in non-commodity GVCs; and (3) measures of innovation (intellectual property or IP receipts as a percentage of GDP and research and development or R&D intensity, defined as its expenditure of public and private R&D as a percentage of GDP). The formulated taxonomy is

Commodities, Limited manufacturing, Advanced manufacturing and services, and Innovative activities. The "Commodities" is defined as the manufacturing share of total domestic value added in exports is less than 60 per cent with backward manufacturing less than 20 per cent, 10 per cent, and 7.5 per cent for small, medium, and large countries respectively. These criteria ensure that manufacturing is a small share of exports and that backward linkages in manufacturing are limited. This group is further divided into low participation, limited commodities, and high commodities based on less than 20 per cent, equal to or greater than 20 per cent but less than 40 per cent, and equal to or greater than 40 per cent respectively in primary goods' share of total domestic value added in exports. These criteria define countries according to their export dependence on manufacturing. 'Innovative activities' (based on remaining countries) is defined for small countries as equal to or greater than 0.15 per cent in IP receipts & 1.5 per cent R&D expenditure of GDP and for medium & large countries as equal to or greater than 0.1 per cent in IP receipts & 1 per cent R&D expenditure of GDP. These criteria split groups into those that spend a relatively large share of GDP on research and receive a large share of GDP from intellectual property. 'Advanced manufacturing and services' (based on remaining countries) are defined as the share of manufacturing and business services in total domestic value added in exports is equal to or greater than 80 per cent. It is again categorised for small, medium, and large countries based on backward manufacturing is equal to or greater than 30 per cent, 20 per cent, and 15 per cent respectively. Here business services include maintenance and repair; wholesale trade; retail trade; transport; post and telecommunications; and financial intermediation and business activities. Business services rather than total services were used to detect advanced countries with a developed services sector. Lastly, the 'limited manufacturing' is defined on the rest of the sample.

It looks like a ladder of GVC upgradation where on average Commodities have the highest scope for the forward participation due to being used in various downstream production processes. Limited manufacturing has lower forward participation than the backward participation, and Advanced manufacturing & services have the highest scope for backward participation. Innovative activities in this sense are exceptional with slightly lower backward participation as these are less reliant on imported inputs.

**GVC Participation across Regions, Countries, Sectors, and Firms**

Mainly developed countries of North America, Western Europe, and East Asia dominate in GVCs of advanced manufacturing and services, whereas Africa, Central Asia, and Latin America are mostly in commodities and limited manufacturing GVCs. India belongs to the Advanced manufacturing & services category. Between 1990 and 2015, many countries transitioned up into more sophisticated GVCs. It is also interesting that South Asia, Latin America, Middle East, and Africa regions increased globally in GVC activities. Whereas, Europe and East Asia regions expanded the GVC activities regionally. Indeed, a handful of countries drive the GVC expansion. Though developed countries (Germany, USA, Japan, Italy, and France) intensified GVC growth using imported components in their exports, China significantly scaled up GVC growth in terms of its share of international trade.

The sectoral composition of GVC participation is very much diverse. Some sectors have contributed heavily to GVCs for decades. Although on average the essential sectors with intensive use of resources and imported inputs (chemicals, petroleum, metals, rubber, and plastics) have increased GVC participation over time, textiles and leather have decreased to some extent. In services, construction and transport-related activities have performed well. High-tech manufacturing sectors (downstream industries like electrical and optical equipment, transport equipment, and machinery) intensified GVC through increased use of imported inputs. Whereas, low-tech manufacturing sectors (upstream industries like mining and coke) scaled up GVC in terms of a growing share of international trade over the period. GVCs have also increased rapidly in the service sector. As GVC is about trade in tasks, MNCs also outsource service tasks. These services (transportation, telecommunication, and financial services) act as facilitation and coordination of geographically fragmented production process across sectors. The developed countries (France, Germany, Italy, UK, and USA) contribute more than half of value-added exports. India's expanding GVC role in ICT and Business services is genuinely remarkable.

As firms, not countries or industries participate in world trade, the last two decades of economic research has seen the importance of data availability and theoretical transformation regarding firm-level trade studies. In this literature, firm heterogeneity in terms of productivity leads to export & import decisions. It implies to the stylised fact that on average, a small share (15 per cent of all trading firms) of big import-export firms (two-way traders) dominate international trade (80 per cent of total trade) across countries. Therefore, they dominate GVC participation as 'superstar' or 'lead' firms setting up networks of upstream and downstream economic activities. Nevertheless, intensive measures of firm-level GVC participation are challenging.

Besides this, the emergence of this firm-level GVCs under sunk costs, customisation, limited contractual security leads to stickiness among firms relating to matching buyers and sellers, relationship-specific investments, flows of intangibles. In other words, the identity of the economic agents participating in a GVC is crucial, and within GVCs relationships are more likely to show persistence. These relational aspects of the growth of GVCs also exemplify intrafirm trade flows. At the global level, intrafirm trade has contributed about one third of world trade flows. With the firm-level approach, one can also distinguish between 'producer-driven' (like Apple) and 'buyer-driven' (like Walmart) GVCs based on the complexity of products, the ability to codify transactions, and the capabilities of supply firms.

Generally, MNCs control these relational contracting concerning market-seeking investment and efficiency-seeking investment. Both forces of the globalised production process are crucial for the alignment between GVC participation and FDI flows. Though there is a positive correlation between FDI inflows and GVC participation in both high-income and low- & middle-income countries, FDI outflows are relatively high for high-income countries with GVC growth.